# Gravitational waves from binary supermassive black holes missing in pulsar observations

**Authors:** R. M. Shannon[1,2][*], V. Ravi[3][*], L. T. Lentati[4], P. D. Lasky[5], G. Hobbs[1], M. Kerr[1], R. N. Manchester[1], W. A. Coles[6], Y. Levin[5], M. Bailes[3], N. D. R. Bhat[2], S. Burke-Spolaor[7], S. Dai[1,8], M. J. Keith[9], S. Osłowski[10,11], D. J. Reardon[5], W. van Straten[3], L. Toomey[1], J.-B. Wang[12], L. Wen[13], J. S. B. Wyithe[14], X.-J. Zhu[13]

**Affiliations:**

[1] CSIRO Astronomy and Space Science, Australia Telescope National Facility, P.O. Box 76, Epping, NSW 1710, Australia.

[2] International Centre for Radio Astronomy Research, Curtin University, Bentley, WA 6102, Australia.

[3] Centre for Astrophysics and Supercomputing, Swinburne University of Technology, PO Box 218, Hawthorn, VIC 3122, Australia.

[4] Astrophysics Group, Cavendish Laboratory, JJ Thomson Avenue, Cambridge CB3 0HE, UK.

[5] Monash Centre for Astrophysics, School of Physics and Astronomy, Monash University, PO Box 27, VIC 3800, Australia.

[6] Department of Electrical and Computer Engineering, University of California at San Diego, La Jolla, CA 92093, USA.

[7] National Radio Astronomical Observatory, Array Operations Center, P.O. Box O, Socorro, NM 87801-0387, USA.

[8] Department of Astronomy, School of Physics, Peking University, Beijing, 100871, China.

[9] Jodrell Bank Centre for Astrophysics, University of Manchester, M13 9PL, UK.

[10] Department of Physics, Universitat Bielefeld, Universitatsstr 25, D-33615 Bielefeld, Germany.

[11] Max-Planck-Institut für Radioastronomie, Auf dem Hügel 69, 53121 Bonn, Germany.

[12] Xinjiang Astronomical Observatory, CAS, 150 Science 1-Street, Urumqi, Xinjiang 830011, China.

[13] School of Physics, University of Western Australia, Crawley, WA 6009, Australia.

[14] School of Physics, University of Melbourne, Parkville, VIC 3010, Australia.

*Correspondence to: Ryan Shannon (ryan.shannon@csiro.au), Vikram Ravi (v.vikram.ravi@gmail.com)

**Abstract**: Gravitational waves are expected to be radiated by supermassive black hole binaries formed during galaxy mergers. A stochastic superposition of gravitational waves from all such binary systems will modulate the arrival times of pulses from radio pulsars. Using observations of millisecond pulsars obtained with the Parkes radio telescope, we constrain the characteristic amplitude of this background, $A_{c,yr}$, to be $< 1.0 \times 10^{-15}$ with 95% confidence. This limit excludes predicted ranges for $A_{c,yr}$ from current models with 91-99.7% probability. We conclude that binary evolution is either stalled or dramatically accelerated by galactic-center environments, and that higher-cadence and shorter-wavelength observations would result in an increased sensitivity to gravitational waves.

Studies of the dynamics of stars and gas in nearby galaxies provide strong evidence for the ubiquity of supermassive (> $10^6$ solar mass) black holes (SMBHs) (*1*). Observations of luminous quasars indicate that SMBHs are hosted by galaxies throughout the history of the universe (*2*) and affect global properties of the host galaxies (*3*). The prevailing dark energy – cold dark matter cosmological paradigm predicts that large galaxies are assembled through the hierarchical merging of smaller galaxies. The remnants of mergers can host gravitationally bound binary SMBHs with orbits decaying through the emission of gravitational waves (GWs) (*4*).

Gravitational waves from binary SMBHs, with periods between ~ 0.1 and 30 yr (*5*), can be detected or constrained by monitoring, for years to decades, a set of rapidly rotating millisecond pulsars (MSPs) distributed throughout our galaxy. Radio emission beams from MSPs are observed as pulses that can be time-tagged with as small as 20 ns precision (*6*). When traveling across the pulsar-Earth line of sight, GWs induce variations in the arrival times of the pulses (*7*).

The superposition of GWs from the binary SMBH population is a stochastic background (GWB), which is typically characterized by the strain-amplitude spectrum $h_c(f) = A_{c,yr}[f/(1\ yr^{-1})]^{-2/3}$, where $f$ is the GW frequency, $A_{c,yr}$ is the characteristic amplitude of the GWB measured at $f = 1\ yr^{-1}$, predicted to be $A_{c,yr} > 10^{-15}$ (*5, 8-12*), and –2/3 is the predicted spectral index (*5, 8-12*). The GWB will add low-frequency perturbations to pulse arrival times. While the detection of the GWB would confirm the presence of a cosmological population of binary SMBHs, limits on its amplitude constrain models of galaxy and SMBH evolution (*8*).

As part of the Parkes Pulsar Timing Array project to detect GWs (*6*), we have been monitoring 24 pulsars with the 64-m Parkes radio telescope. We have produced a new data set, using observations taken at a central wavelength of 10 cm and previously reported methods (*6,8*), that spans 11 yr, which is 3 yr longer than previous data sets analyzed at this wavelength. In addition to having greater sensitivity to the GWB because of the longer duration, the data set was improved by identifying and correcting for some instrumental offsets (see supplementary section S1, *13*).

We searched for a GWB in observations of the four pulsars (Fig. 1) that have the highest timing precision and therefore are most sensitive to the GWB. Observations of these pulsars at other wavelengths contain excess noise inconsistent with the 10 cm observations, and were therefore excluded from this analysis (see supplementary section S2.1, *13*). This does not bias our analysis because GWs produce achromatic variations in arrival times. Observations of other pulsars are not presented here because they have insufficient timing precision, relative to the best pulsars, to influence the search (see supplementary section S2, *13*). We also have not corrected for chromatic arrival-time variations associated with propagation through a varying column of interstellar plasma, because these effects are small in the 10 cm band (*14*). Additionally, using uncorrected observations can only reduce our sensitivity to the GWB, making our analysis conservative.

We used a Bayesian methodology (*15*) to marginalize over the pulsar rotational ephemerides and search for stochastic contributions to the arrival times. The stochastic terms include excess white noise associated with intrinsic pulse-shape changes and instrumental distortions uncorrelated between observations. They also include excess low-frequency timing noise that is uncorrelated between pulsars, which could be intrinsic to the pulsars or caused by interstellar propagation effects. Finally, the model includes the GWB, which produces timing perturbations that are correlated between the pulsars (*7*). The methodology also enables us to quantitatively compare models by providing evidence, in the form of a probability, which can be used to select a preferred model (see supplementary section S2, *13*).

We find no evidence for the GWB in our data set. We therefore place an upper limit on the amplitude of the GWB by analyzing its posterior distribution. The pulsar that individually provides the best limit on the GWB, PSR J1909–3744 (Fig. 1), shows no evidence for excess low-frequency timing noise. The pulsar that provides the third most constraining limit, PSR J0437-4715 (Fig. 1), shows evidence for a low-frequency signal that is inconsistent both with the predicted GWB spectral shape (99.0% probability), and in amplitude (99.4% probability) with the limit derived from PSR J1909–3744.

For the GWB strain-amplitude spectrum $h_c(f) = A_{c,yr}[f/(1\ yr^{-1})]^{-2/3}$, we find $A_{c,yr} < 1.0 \times 10^{-15}$ with 95% probability (Fig. 2). As a fraction of the critical density of the universe, per logarithmic GW frequency, the limit is $\Omega_{GW} < 2.3 \times 10^{-10}$ at a frequency of 0.2 yr$^{-1}$ (see supplementary section S2.2, *13*). This is a factor of six lower than any previous limit (Fig. 2). Other PTA experiments with comparable data spans, but observe at longer wavelengths, do not achieve the same sensitivity to GWs (*16*) because of the higher timing precision in our observations, and the presence of low frequency noise in theirs. Our limit is inconsistent with current models for the GWB (*9-12*) with between 91% and 99.7% probability (see supplementary section S3, *13*).

Our limit on the GWB therefore suggests that at least one of the physical assumptions underlying these models is incorrect. Models for the binary SMBH population rely on measurements of the galaxy merger rate. They also assume that all galaxy mergers form binary SMBHs that coalesce well before a subsequent galaxy merger, and that binary orbital decay is driven only by losses of energy to GWs when radiating in the pulsar-timing frequency band (Fig. 3, black curve).

Fig. 3 displays schematically the evolution of a binary SMBH in which each component has a mass of 10$^9$ solar masses, evolving under standard assumptions (Fig. 3, blue curve) and in other ways that produce a weaker GWB.

Larger galaxy merger timescales would result in a lower inferred merger rate (Fig. 3, green curve), fewer binary SMBHs, and hence a lower GWB amplitude. While predictions for merger timescales vary by a factor of three (*17*), models that include this uncertainty (*9,11*) are in tension with our limit. Shorter predicted timescales result from the inclusion of more sophisticated physical mechanisms and are favored (*17*). Therefore, galaxy mergers are expected to rapidly form gravitationally bound binaries.

Models for the GWB also assume that all large galaxies host SMBHs. A low SMBH occupation fraction beyond the local Universe ($z > 0.3$) could result from exceedingly rare high-redshift SMBH seed formation (*18*). For this to be the case, seed SMBHs would have to occupy ~1% of the most massive galaxies at z ~ 6 (*19*). The models also assume that, post-coalescence, SMBHs remain gravitationally bound to their host galaxies. However, it is unlikely that the acceleration of post-coalescence SMBHs beyond galactic escape velocities through gravitational-radiation recoil (Fig. 3, purple-dashed curve) results in a significant number of galaxies being devoid of SMBHs (*20*).

The GWB amplitude would also be reduced if SMBH binaries do not efficiently reach the GW-emitting stage (Fig. 3, red curve). Dynamical friction is expected to bring the SMBHs in a merging galaxy pair close enough to form a bound binary (*4*), with an orbital major axis $a_{\rm form} \sim 60\, M_9^{0.54}$ pc, where $M_9$ is the mass of the larger SMBH in units of $10^9$ solar masses (see supplementary section S4, *13*). The time to coalescence through GW emission is $t_{\rm gw} = 18\, M_9^{-3}[a_{\rm form}/(1\, {\rm pc})]^4$ Gyr in the lower-limiting case of an equal-mass binary, which is longer than the age of the universe. Hence another mechanism, besides GW emission, is required to drive binaries to coalescence.

Observations and theoretical models, however, indicate that binary SMBHs can coalesce within the age of the universe through the coupling of binary SMBHs to their environments (*21*). Proposed coupling mechanisms include the three-body scattering of stars on radial orbits and viscous friction against circumbinary gas. The action of environments (Fig. 3, gray curve) would cause binaries to spend less time emitting GWs, reducing the GWB amplitude at low frequencies. Our non-detection of the GWB may therefore result from the efficient coupling of binary SMBHs to their environments (*10,22,23*).

Modeling of the stellar environments of the cosmological population of binary SMBHs (*22*) indicates that the GWB characteristic-strain spectrum may be attenuated at frequencies up to 0.3 yr$^{-1}$ (Fig. 2); similar results are obtained when the possible gas-rich environments of binary SMBHs are considered (*23*). Our GWB constraint, placed at 0.2 yr$^{-1}$, is consistent with some models that predict the extreme efficiency of environments in shrinking SMBH binary orbits (Model R14, Fig. 2). However other environmentally driven models that include higher galaxy merger rates (*10*) are inconsistent with our limit (Model Exp, Fig. 2)

Distinguishing between explanations for our limit requires further observations and better models of SMBH evolution. The characterization of a substantial population of binary or recoiling SMBHs (*24*) would better delineate the coalescence rate. The coalescence events themselves may produce strong millihertz-frequency GWs that could be detected by space-based laser interferometers (*5*). The detection of the GWB at frequencies $\gtrsim 0.2$ yr$^{-1}$ with the currently predicted amplitude would provide strong evidence for the high efficiency of binary environments in shrinking orbits (*25*). This hypothesis also predicts an enhanced prospect for detecting low-frequency GWs from the most massive individual binary SMBHs, which are less affected by their environments (*22,23*). The alternate explanation for our limit is that the SMBH-SMBH coalescence rate is lower than current estimates suggest; in this case the GWB may still have a power-law spectrum.

This limit implies that a change in observational strategy could increase the sensitivity of pulsar timing arrays to gravitational waves. One approach is to obtain more observations of pulsars with comparable sensitivity to those of our four best pulsars. If the observed excess noise at longer radio wavelengths is astrophysical, observations will need to be conducted at shorter wavelengths ($\lesssim 10$ cm). In this case, GWB detection may require observations with a sensitive radio telescope such as the Square Kilometre Array (SKA, Ref. *26*), because MSP emission is weaker at these wavelengths. If binary SMBH environments are driving orbital evolution, a high-cadence campaign is required to detect the GWB at frequencies $\gtrsim 0.2$ yr$^{-1}$. Alternatively, the GWB could have a power-law spectrum but be weak in amplitude. Our limit implies that there is a 50% probability that $A_{\rm c,yr} < 2.4 \times 10^{-16}$. In this case, the first evidence for the GWB will be low-frequency perturbations to timing observations of the most stable pulsar, PSR J1909–3744, when longer data spans are achieved. In any case, the predicted time to detection of the GWB with pulsar timing arrays (*27*) has been underestimated.

It is also possible that there is a more exotic reason for our non-detection. We have not yet tested GWBs expected from alternate theories of gravity. Our limit is consistent with GWs being absorbed on cosmological scales (*28*). Until GWs are detected, our limits will continue to improve with data span, as more pulsars are added into the sample, and improved analysis methods are developed (Fig. 2, blue pentagon). These limits will provide even stronger constraints on models of supermassive black hole formation and evolution.

**References and Notes:**

1. J. Kormendy, L. Ho, Coevolution (or not) of supermassive black holes and host galaxies. *Annu. Rev. Astron. Astrophys.* **51**, 511 (2013).
2. F. Shankar, D. H. Weinberg, J. Miralda-Escudé, Accretion-driven evolution of black holes: Eddington ratios, duty cycles and active galaxy fractions. *Mon. Not. R. Astron. Soc.* **428**, 421 (2013).
3. A. Fabian, Observational Evidence of Active Galactic Nuclei Feedback. *Annu. Rev. Astron. Astrophys.* **50**, 455 (2012).
4. M. Begelman, R. Blandford, M. Rees, Massive black hole binaries in active galactic nuclei. *Nature* **287**, 307 (1980).
5. J. S. B Wyithe, A Loeb, Low-Frequency Gravitational Waves from Massive Black Hole Binaries: Predictions for LISA and Pulsar Timing Arrays, *Astrophys. J.,* **590**, 691 (2003).
6. R. N. Manchester *et al.*, The Parkes Pulsar Timing Array project. *Publ. Astron. Soc. Aust.* **30**, e017 (2013), doi:10.1017/pasa.2012.017.
7. R. W. Hellings, G. S. Downs, Upper limits on the isotropic gravitational radiation background from pulsar timing analysis. *Astrophys. J.* **265**, L39 (1983).
8. R. M. Shannon *et al.*, Gravitational-wave limits from pulsar timing constrain supermassive black hole evolution. *Science* **342**, 334 (2013).
9. A. Sesana, Insights into the astrophysics of supermassive black hole binaries from pulsar timing observations. *Classical Quant. Grav.* **30**, 224014 (2013), doi:10.1088/0264-9381/30/22/224014.
10. S. T. McWilliams, J. P. Ostriker, F. Pretorius, Gravitational waves and stalled satellites from massive galaxy mergers at z <= 1. *Astrophys. J.* **789**, 156 (2014), doi:10.1088/0004-637X/789/2/156.
11. V. Ravi, J. S. B. Wyithe, R. M. Shannon, G. Hobbs, Prospects for gravitational-wave detection and supermassive black hole astrophysics with pulsar timing arrays. *Mon. Not. R. Astron. Soc.* **447**, 2772-2783 (2015).
12. A. Kulier, J. P. Ostriker, P. Natarajan, C. N. Lackner, R. Cen, Understanding black hole mass assembly via accretion and mergers at late times in cosmological simulations. *Astrophys. J.*, **799,** 178 (2015)**,** doi:10.1088/0004-637X/799/2/178
13. Materials and methods are available as supporting materials on Science Online.
14. M. J. Keith *et al.*, Measurement and correction of variations in interstellar dispersion in high-precision pulsar timing. *Mon. Not. R. Astron. Soc.* **429**, 2161 (2013).
15. L. Lentati *et al.*, Hyper-efficient model-independent Bayesian method for the analysis of pulsar timing data. *Phys Rev. D.* **87,** 104021 (2013), doi:10.1103/PhysRevD.87.104021

16. L. Lentati *et al.*, European Pulsar Timing Array limits on an isotropic stochastic gravitational-wave Background. *Mon. Not. R. Astron. Soc.*, submitted. (available at http://arxiv.org/abs/1504.03692).
17. C. Conselice, The evolution of galaxy structure over cosmic time. *Annu. Rev. Astron. Astrophys.* **52**, 291 (2014).
18. K. Menou, Z. Haiman, V. K. Narayanan, The merger history of supermassive black holes in galaxies. *Astrophys. J.* **558**, 535 (2001).
19. T. L. Tanaka Driving the growth of the earliest supermassive black holes with major mergers of host galaxies. *Classical Quant. Grav* **31**, 4005 doi:10.1088/0264-9381/31/24/244005 (2014).
20. Gerosa, A. Sesana, Missing black holes in brightest cluster galaxies as evidence for the occurrence of superkicks in nature. *Mon. Not. R. Astron. Soc.* **446**, 38 (2015).
21. M. Colpi, Massive binary black holes in galactic nuclei and their path to coalescence. *Space Sci. Rev.* **183**, 189 (2014).
22. V. Ravi, J. S. B. Wyithe, R. M. Shannon, G. Hobbs, R. N. Manchester, Binary supermassive black hole environments diminish the gravitational wave signal in the pulsar timing band. *Mon. Not. R. Astron. Soc.* **442**, 56 (2014).
23. B. Kocsis, A. Sesana, Gas-driven massive black hole binaries: signatures in the nHz gravitational wave background. *Mon. Not. R. Astron. Soc.* **411,** 1467 (2011).
24. M. Eracleous, T. A. Boroson, J. P. Halpern, J. Liu, A Large Systematic Search for Close Supermassive Binary and Rapidly Recoiling Black Holes. *Astrophys. J. Supp.* **201**, 23, doi: 10.1088/0264-9381/31/24/244005 (2012).
25. L. Sampson, N. J. Cornish, S. T. McWilliams, Constraining the solution to the last parsec problem with pulsar timing. *Phys. Rev. D* **91**, 084055 (2015).
26. G. H. Janssen *et al.*, Gravitational wave astronomy with the SKA. *Proc. Sci.*, in press (available at http://arxiv.org/abs/1501.00127).
27. X. Siemens, J. Ellis, F. Jenet, J. D. Romano, The stochastic background: scaling laws and time to detection for pulsar timing arrays. *Classical Quant. Grav.* **30**, 224015, doi:10.1088/0264-9381/30/22/224015 (2013).
28. S. W. Hawking, Perturbations of an expanding universe. *Astrophys. J.*, **145**, 544 (1966).
29. P. B. Demorest, *et al.* Limits on the Stochastic Gravitational Wave Background from the North American Nanohertz Observatory for Gravitational Waves. *Astrophys. J.* **762**, 94. (2013)
30. http://www.bipm.org/en/bipm/tai/
31. R. van Haasteren *et al.*, On measuring the gravitational-wave background using Pulsar Timing Arrays. *Mon. Not. R. Astron. Soc.* **395**, 1005 (2009).
32. R. van Haasteren, Y. Levin, Understanding and analysing time-correlated stochastic signals in pulsar timing. *Mon. Not. R. Astron. Soc.* **428**, 1147 (2013).
33. W. van Straten, High-fidelity radio astronomical polarimetry using a millisecond pulsar as a polarized reference source. *Astrophys. J. Supp.* **204**, 13 (2013), doi:10.1088/0067-0049/204/1/13.
34. J. M. Cordes, R. M. Shannon, A measurement model for precision pulsar timing. Available at http://arxiv.org/abs/1010.37852010 (2010).
35. R. M. Shannon *et al.*, Limitations in timing precision due to single-pulse shape variability in millisecond pulsars. *Mon. Not. R. Astron. Soc.* **443**, 1463 (2014).

36. J. H. Taylor, Pulsar timing and general relativity. *Phil. Trans. Phys. Sci. Eng.* **341**, 117 (1992).
37. R. M. Shannon, J. M. Cordes, Assessing the Role of Spin Noise in the Precision Timing of Millisecond Pulsars. *Astrophys. J.* **725,** 1607 (2010).
38. F. Feroz, M. P. Hobson, M. Bridges, MultiNest: an efficient and robust Bayesian inference tool for cosmology and particle physics. *Mon. Not. R. Astron. Soc.* **398**, 1601 (2009).
39. L. Lentati *et al*., TEMPONEST: a Bayesian approach to pulsar timing analysis. *Mon. Not. R. Astron. Soc.* **437**, 3004 (2014).
40. R. van Haasteren, M. Vallisneri, Low-rank approximations for large stationary covariance matrices, as used in the Bayesian and generalized-least-squares analysis of pulsar-timing data. *Mon. Not. R. Astron. Soc.* **446**, 1170 (2015).
41. A. Melatos, B. Link, Pulsar timing noise from superfluid turbulence. *Mon. Not. R. Astron. Soc.* **437**, 21 (2014).
42. W. W. Zhu *et al.*, Testing Theories of Gravitation Using 21-Year Timing of Pulsar Binary J1713+0747. *Astrophys, J.*, submitted. (Available at http://arxiv.org/abs/1504.00662)
43. Z. Arzoumanian *et al.*, The NANOGrav Nine-year Data Set: Observations, Arrival Time Measurements, and Analysis of 37 Millisecond Pulsars. *Astrophys, J.*, submitted. (Available at http://arxiv.org/abs/1505.07540)
44. Z. Arzoumanian *et al.*, Gravitational Waves from Individual Supermassive Black Hole Binaries in Circular Orbits: Limits from the North American Nanohertz Observatory for Gravitational Waves. *Astrophys, J.*, **794**, 141 (2014), doi:10.1088/0004-637X/794/2/141
45. G. Hobbs *et al.,* TEMPO2: a new pulsar timing package - III. Gravitational wave simulation. *Mon. Not. R. Astron. Soc.* **394**,1945 (2009).
46. K. J. Lee *et al*., Model-based asymptotically optical dispersion measure correction for pulsar timing. *Mon. Not. R. Astron. Soc.* **441**, 2831 (2014).
47. Q. Yu, Evolution of massive binary black holes. *Mon. Not. R. Astron. Soc.* **331**, 935-958 (2002).
48. G. D. Quinlan, The dynamical evolution of massive black hole binaries I. Hardening in a fixed stellar background. *New Astron.* **1**, 35 (1996).
49. A. Escala, R. B. Larson, P. S. Coppi, D. Mardones, The role of gas in the merging of massive black holes in galactic nuclei. I. black hole merging in a spherical gas cloud. *Astrophys. J.* **605**, 765 (2004).
50. L. Mayer *et al.*, Rapid formation of supermassive black hole binaries in galaxy mergers with gas. *Science* **316**, 1874 (2007).
51. S. Chandrasekhar, Dynamical friction. I. General considerations: the coefficient of dynamical friction. *Astrophys. J.* **97**, 255 (1943).
52. A. Sesana, Self consistent model for the evolution of eccentric massive black hole binaries in stellar environments: implications for gravitational wave observations. *Astrophys. J.* **719**, 851 (2010).
53. J. A. Gonzalez *et al.*, Supermassive Recoil Velocities for Binary Black-Hole Mergers with Antialigned Spins. *Phys. Rev. Lett.* **98**, 231101 (2007).
54. M. Campanelli, C. Lousto, Y. Zlochower, D. Merritt, Large Merger Recoils and Spin Flips from Generic Black Hole Binaries. *Astrophys. J.* **659**, L5 (2007).

**Acknowledgments:** We thank all of the observers, engineers, and Parkes observatory staff members who have assisted with the observations reported in this paper. We thank R. van Haasteren for assistance with the use of the code *piccard*, E. Thomas for comments on the manuscript, and I. Mandel for discussions on model selection. The Parkes radio telescope is part of the Australia Telescope National Facility which is funded by the Commonwealth of Australia for operation as a National Facility managed by Commonwealth Science and Industrial Research Organization (CSIRO). The Parkes Pulsar Timing Array Project project was initiated with support from RNM's Australian Research Council (ARC) Federation Fellowship (FF0348478) and from the CSIRO under that fellowship program. The PPTA project has also received support from ARC through Discovery Project grants DP0985272 and DP140102578. NDRB acknowledge support form a Curtin University research fellowship. GH and YL are recipients of ARC Future Fellowships (respectively, FT120100595 and FT110100384). SO is supported by the Alexander von Humboldt Foundation. RMS acknowledges travel support from the CSIRO through a John Philip award for excellence in early career research. The authors declare no conflicts of interest. Data used in this analysis can be accessed via the Australian National Data Service (http://www.ands.org.au).

SUPPLEMENTARY MATERIALS

Supplementary Text
Figs. S1 to S2
Tables S1 to S8

References (*30-54*)

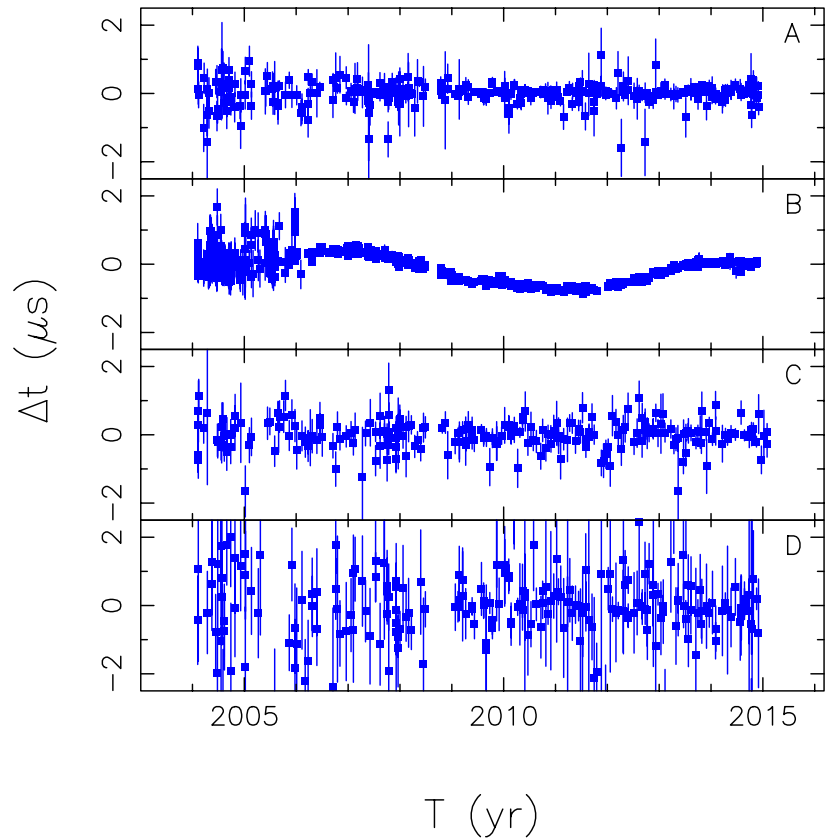

**Fig. 1. Residual pulse times of arrival, Δ*t*, for the four pulsars used in our analysis.** These are PSR J1909-3744 (panel *A*), PSR J0437-4715 (panel *B*), PSR J1713+0747 (panel *C*), and PSR J1744-1134 (panel *D*).

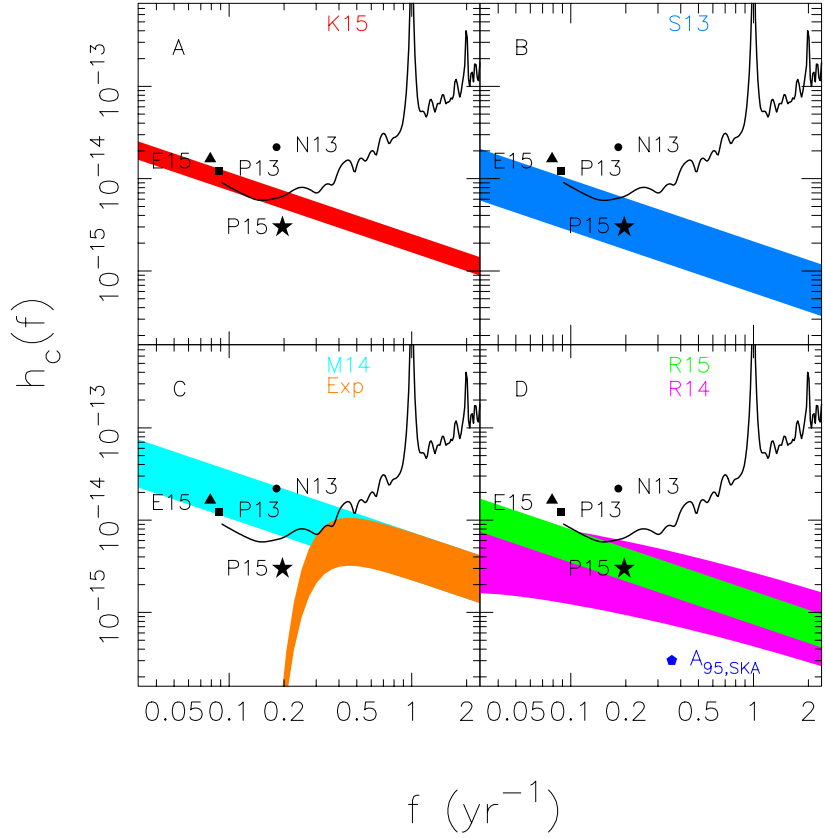

**Fig. 2. Predictions and limits on the GWB strain spectrum.** The black asterisks (labeled P15) shows the 95% confidence limit we obtain, assuming $h_c(f)=A_{c,yr}[f/(1\ yr^{-1})]^{-2/3}$. The other symbols show previously published limits from the European Pulsar Timing Array (triangle, labeled E15, Ref. *20*), the North American Nanohertz Observatory for Gravitational Waves (circle, labeled N13, Ref. *29*) collaborations, and our previous limit (square, labeled P13, Ref. *8*). Each panel shows a different prediction for the GWB as a shaded region that represents the 1-$\sigma$ uncertainty, including four models for SMBH evolution, labeled S13 (*9*), M14 (*10*), K15 (*12*), and R15 (*11*), which predict a power-law form for $h_c(f)$. Models Exp (See supplementary section S2.2, Ref. *13*) and R14 (*22*) include the effects of environmentally driven binary evolution and therefore predict more complex strain spectra. The black curves show the nominal single-frequency sensitivities of our observations (see supplementary section S2.2, *13*), and is above our limit because of the statistical penalties applied when searching individual frequencies. In Panel *D*, the blue pentagon (labeled $A_{95,SKA}$) shows the projected upper limit on $A_{c,yr}$ obtained with a single-pulsar timing campaign with a next generation radio telescope (the SKA; see supplementary section S2.2, Ref. *13*), and excludes all models considered with greater than 98% probability.

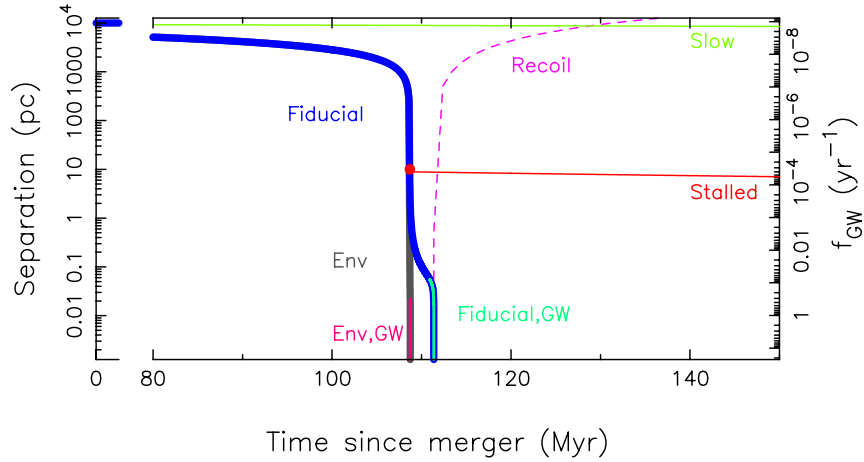

**Fig. 3. Illustrative evolutionary paths for a pair of $10^9$ solar-mass SMBHs in a galaxy merger.** The figure shows the pair separation and the GW emission frequency $f_{GW}$, assuming the binary is in a circular orbit. The blue curve shows the evolution of the separation of the SMBHs using fiducial assumptions, which results in a GWB that is inconsistent with our data. The cyan curve labeled *Fiducial,GW* is the portion of the evolution when GW-emission dominates orbital decay. We also show scenarios that could explain our GWB limit. First, the galaxy merger rate could be lower, as represented by the *slow* merger curve (green curve). Alternatively, after the SMBHs form a binary (red circle), the orbital evolution may stall prior before emitting GWs (red curve). The gray curve shows a scenario in which a dense binary SMBH environment drives orbital decay through the GW frequency band at which we are sensitive. In this case, GW emission dominates only for $f_{GW} > 0.5$ yr$^{-1}$ (pink curve, labeled *Env,GW*). Finally, it is possible that the post-coalescence SMBH could undergo gravitational recoil and escape its host galaxy (purple dashed curve), negating the possibility of it again forming a binary SMBH.

Supplementary Text

**S1. Data set**

We analyzed observations obtained with the 64-m Parkes Telescope as part of the Parkes Pulsar Timing Array (PPTA) project (*6*). Our data set spans approximately 11 years, from MJD 53041 to 56992, which is approximately 3.5 years longer than a previously analyzed multi-wavelength data set (*8*). All of the observations reported here were made in the 10 cm band of the 10-cm/50-cm receiver, at a central frequency of 3100 MHz, with bandwidths of 512 or 1024 MHz and individual durations of usually 3840 s. The observations were recorded with a set of autocorrelation and digital-filterbank spectrometers. The data were analyzed using previously described procedures (*6*). Observations were referred to the solar-system barycentre using the DE421 solar-system ephemeris and the 2014 realization of terrestrial time published by BIPM (*30*). We used observations of four pulsars that had the highest timing precision: PSRs J0437–4715, J1713+0474, J1744–1134, and J1909–3744. Inclusion of other PPTA pulsars does not change the limit, and their exclusion does not bias the results, because they are not sensitive to a GWB of the amplitude at the level bounded by these four pulsars. While we record data at other frequencies (*6*) we do not include it in this analysis, as justified in section S2.2.

Discrete phase offsets in pulse arrival times (referred to as jumps) can occur when changes in the signal-processing instrumentation (referred to as backends) are made which are the result of either hardware (e.g., cable length) or digital signal processing changes (e.g., changes to the digital-filterbank firmware). Previously, these offsets were estimated independently from the pulsar timing observations (*6*) by injecting a pulsed signal into the telescope receiver and measuring the delays between backend systems. For the best pulsars, the offsets were further modified by shifting the analytic standards to compensate for apparent inaccuracies in the *a priori* delay measurements. The combination of these corrections was found to have insufficient accuracy in the latest dataset. With the additional data, the offsets of the largest jump could be seen "by-eye" in the dataset. Three offsets were included to model data of old backends (WBCORR, PDFB1, PDFB2) relative to the most recent instrument. Two additional jumps were identified at epochs where the most recent backend was used (PDFB4). The first occurred at MJD 55319.8 when an update to the backend software changed the location within the program where the data were time stamped. These jumps are not common between pulsars because different pulsars were observed with different backend software .

We included these offsets as free parameters in the timing model. In Bayesian methodology, these jumps are fully accounted for through analytic marginalization. As verified through simulations discussed in the next section, this does not bias the limit placed on the GWB amplitude.

For PSR J1909–3744, which dominates the limit, we reprocessed the data from raw files as recorded at the telescope, using a second independent pipeline, and found consistent results between the two pipelines.

## S2. Timing analysis and calculating a limit for the GWB amplitude

We used Bayesian methodologies to model the pulse arrival times (*15,31-32*). The timing model includes the deterministic ephemeris of the pulsar and stochastic parameters that describe noise processes in the pulsar. The stochastic parameters include white-noise and red-noise components. The white-noise terms are temporally uncorrelated contributions that model excess noise (beyond the formal TOA uncertainties) and may be associated with instrumental errors such as polarization calibration (*33*) or intrinsic pulse profile instability (*34-35*). We modeled the white noise by modifying the TOA uncertainties through the transformation $\sigma_F^2 = F\sigma_M^2 + Q^2$, where $\sigma_M$ is the uncertainty derived from the profile-fitting used to measure the TOAs (*36*), and *F* and *Q* (referred to respectively as *EFAC* and *EQUAD*) are the white-noise parameters. The red-noise terms are time-correlated contributions (with excess power at low frequencies) and include both a GWB and power-law red noise of unknown spectral index that could be associated with, for example, intrinsic rotational instabilities (*37*). The red noise was assumed to have a power law form of $P_r(f) = M_r(f/1 \text{ yr}^{-1})^\beta$, where $M_r$ is the amplitude of the noise and $\beta$ is its spectral index, while the GWB, unless otherwise stated, is assumed to have the form $P_r(f) = (A_{c,yr}^2/12\pi^2)(f/1 \text{ yr}^{-1})^{-13/3}$.

We marginalized analytically over the deterministic spin ephemeris of the pulsar. This ephemeris includes the pulsar spin frequency and its first derivative, the relativistic orbit of the pulsar about its companion (if the pulsar is in a binary system, as is the case for three of the pulsars in our sample), and instrumental offsets between the backends. We used the *multinest* algorithm (*38*) as implemented in *temponest* (*39*), as well as a Markov-Chain Monte Carlo method as implemented in the code *piccard* (*40*), to sample the likelihood of the stochastic parameters. The maximum-likelihood residuals for all four pulsars are displayed in Fig 1. A 95% limit was determined from the posterior distribution of $A_{c,yr}$. To place a limit on the GWB we used a flat prior on $A_{c,yr}$.

We used Bayesian evidence to compare models of stochastic contributions to the TOAs. The inclusion of additional parameters is supported if the evidence increases substantially, which is usually defined as an increase in the logarithm of the evidence of greater than 3 (which represents a probability of 95%). The *multinest* approach to parameter estimation (*15,38-39*) directly provides this evidence. In Table S1, we show the evidence for models that contain excess white noise, excess red noise, and a GWB for individual pulsars in the sample.

For PSRs J1909–3744 and J1744–1134, we find no evidence for red noise or a GWB in the data set. For PSR J1713+0747, we find evidence for red noise but none for a GWB of the assumed form. The red noise has a shallow spectrum (the spectral index is $\beta \approx -1$) and is inconsistent in amplitude with limits on red noise observed (with 99% probability) in PSR J1909–3744, so it is unlikely to be a GWB, or evidence for a GWB of a different form to that considered here. We defer the results of searches for GWBs with different spectral forms to a future work.

For PSR J0437–4715, we find evidence for a steeper red spectrum, with a spectral index that is marginally consistent with a GWB. However, when we compare models that explain the noise as being associated with a GWB (with a fixed power law spectrum with $\beta = -13/3$), to those that search for the spectral index of the power law, we find that, despite the penalty associated with searching over a larger parameter space, the former is disfavoured in evidence by a factor of $e^{-4.5}$, which is 99.0% probability. Additionally, if we assume that the red noise in PSR J0437–4715 is the result of a GWB, we find that the inferred posterior distribution is inconsistent with 99.4% probability with the limit on $A_{c,yr}$ derived from PSR J1909–3744. The higher level of noise is likely to be intrinsic spin noise. The spectral index is consistent with spin noise observed in young pulsars, and some millisecond pulsars (*37*). Models for timing noise in the pulsar population predict that PSR J0437–4715 predict larger levels of intrinsic noise than for PSRs J1909-3744 and J1713+0747, primarily because of the larger amplitude of the spin frequency derivative (*37,41*).

Our maximum-likelihood pulsar ephemerides, displayed in Tables S2-S5, are consistent with previously published ephemerides obtained at Parkes and with other telescopes. In particular, our ephemeris for PSR J1713+0747 provides consistent spin and astrometric measurements to that derived from 18 yr of timing with the Arecibo and Green Bank telescopes (*42*)**,** after converting units and observing epochs. The absences of red noise in PSRs J1744–1134 and PSR J1909–3744 are also consistent with 9 yr of observations from the Green Bank telescope (*43*)**.** We detect modest levels of red noise in PSR J1713+0747. This is unlikely to be DM noise as it is larger by a factor of 3 than the noise measured in longer wavelength observations extrapolated to the 10 cm band (*14*).

In Table S1, we also list the limits on the amplitude of the GWB derived from individual data sets. We find that the limit placed using PSR J1909–3744 is the same as the joint limit, indicating that it is the dominant pulsar in the sample. When excluding PSR J0437–4715 from the analysis, we set a 95% confidence limit of $A_{c,yr} < 8 \times 10^{-16}$. When including PSR J0437–4715 in the analysis, we set a 95% confidence limit of $A_{c,yr} < 1.0 \times 10^{-15}$. The inclusion or exclusion of PSR J1744–1134, which individually provides the fourth most constraining limit in the 10 cm band, does not significantly change the limit we derive. Because the remaining pulsars in the PPTA sample have poorer timing precision in this band (and hence individually set poorer limits on the strength of the GWB), we conclude that their inclusion does not affect our results. The limit is dominated by the uncorrelated noise in the best pulsars. This is consistent with previous searches for GWs in pulsar timing data sets, which have found that only a few pulsars dominate the limit (*44*) or have only focused on only the dominant pulsars (*15*)

Our limit on $A_{c,yr}$ was also confirmed by using a second Bayesian algorithm (*32,40*) to place a limit. This independent code uses a similar model for the data set, but instead uses Monte-Carlo methods instead of nested sampling for parameter estimation.

The validity of the limit method was also tested through simulations. In one set of simulations, the data sets were identical in cadence and TOA uncertainty to the one we observed in PSR J1909–3744, but also included a GWB of amplitude $A_{c,inj}=1.1 \times 10^{-15}$, close in amplitude to the 95% limit that we set. The GWB was injected using a series of 10000 single gravitational-wave sources that were then added to the TOAs (*45*)**.** Out of 100 realisations of this data set, we found that in only 4 realisations, the 95% limit inferred from the posterior distribution was lower than $A_{c,inj}$, indicating that the limit we placed was consistent with a 95% confidence limit (see Fig. S1). Similarly, we found that the 50th percentile of the posterior distribution is consistent with a 50% confidence limit. To further confirm that our limit was valid, we again conducted simulations, but included larger simulated jumps at the epochs of the measured jumps. In each simulation the jumps were randomly chosen from a uniform distribution with a width 10 times larger than the inferred jumps. In these simulations included a GWB of amplitude $1.0\times10^{-15}$. We again found that the simulations provided a valid limit at the stated confidence with the 95% limit being greater than the injected background in 98 of the 100 simulations.

We also applied a previous simulation-based method to the dataset (*8*). In this method a GW detection statistic ($DS_{obs}$) was formed from power spectral density estimates of the residual TOAs. The power spectral density estimates were calculated after searching for and modelling red noise. $DS_{obs}$ was compared to the DS calculated from simulated datasets ($DS_{sim}$) with identical white-noise characteristics, but containing no red noise and only a power-law GWB of a specified strength $A_{inj}$. The 95% confidence limit is set where 95% of the time $DS_{sim} > DS_{obs}$. With this method, we found a higher upper limit of $1.5\times10^{-15}$. Firstly, the method was designed to set a conservative limit on the GWB because the simulated datasets did not include any red noise. Secondly the method relies on searching for red noise in the power spectra of the residual time series, which is suboptimal in datasets containing modelled offsets. In simulated datasets with fewer jumps the simulation-based method produces upper limits in better agreement with the Bayesian methods. Thirdly the noise modelling allowed for non-power law red noise, which again is sub-optimal when searching for a GWB of a specified shape.

Our limit is better than that derived from a comparable data set produced by the European Pulsar Timing Array (*17*). They analyze observations of 6 pulsars that span 8–18 yr. They find evidence for red noise in their best pulsars. For example, in PSR J1909–3744 the red noise detected in the EPTA data set is inconsistent with 90% confidence with the limit on red noise inferred from our observations (we tested this consistency using the model comparison method described in S3). Our observations also have higher timing precision as we obtain a weighted rms residual of 100 ns compared to 130 ns obtained by the EPTA. The noise observed in the EPTA data is consistent with DM variations in the pulsar (*17*); the lack of short-wavelength observations makes it difficult to distinguish red signals (intrinsic or from GWs) from DM variations.

Compared to our previous analysis (*8*), we have improved the limit by both extending the data set and modelling the instrumental offsets. To compare the relative benefit of both improvements we compare limits obtained individually from PSR J1909– 3744, over data spanning our previous data release (DR1, Ref. *6*) and the current data set, and fixing the offsets at the values previously used (*8*) and fitting for them as part of the timing model. In both cases we analyse 10 cm observations (the previous analysis used a combination of 10 cm and 20cm band observations). Over the DR1 data span, we find that the individual limit from J1909–3744 improves from $A_{c,yr} < 7.0\times10^{-15}$ to $3.5\times10^{-15}$ when modelling the jumps. With the current data set the limit improves from $A_{c,yr} < 4.2\times10^{-15}$ to $1.0\times10^{-15}$ when we model the jumps.

## S2.1 Choice of high-frequency observations

We use data at 10 cm and do not correct for dispersion measure (DM) variations. It is common to conduct observations at multiple radio wavelengths to correct for variations in radio-wave propagation time associated with variation in total electron content along the line of sight, referred to as DM variations. Because the effects of DM variations on pulse arrival times scale with the square of wavelength, the 10 cm observations are not strongly affected by DM variations and we can set a superior non-DM corrected single-frequency limit on the amplitude of the GWB. For the pulsars considered here, DM variations both published by us (*14*) and by other groups (*49*) are subdominant to our timing precision and lower than the limit we derive on the GWB.

The longer wavelength observations were affected by excess noise that, even after modeling DM variations, resulted in a poorer limit on the GWB. There are two possible origins for the excess noise. It could be either instrumental in nature, such as polarization calibration errors (*33*), or associated with interstellar propagation that could not be corrected using standard techniques (*14,34,46*).

Preliminary analysis of the noise indicates that it is strongest in the lowest frequency data, suggesting that it is associated with interstellar propagation. As an example, we use Bayesian methodology to characterize the excess low frequency noise in PSR J1909–3744. In addition to considering the 10 cm data as described above, we included 20 cm observations obtained with the same set of spectrometers, and 50 cm observations obtained with the CPSR2, PDFB3 and CASPSR backends (*6*) that spanned the same MJD range as our 10 cm observations.

We search for red noise that is associated with DM variations, achromatic red noise that is common to all three bands (such noise could be associated with intrinsic spin noise or a GWB), 'band noise' that is common to all backend systems within a band but different between bands, and 'system noise' that is common to only one backend instrument. In all cases, we model the noise to be a power law process with unknown amplitude and spectral index. This method will be explored further in subsequent papers.

In the three-band analysis, we find that there is strong evidence (probability > 98%) for excess noise in the 50cm band (see Table S1). There is insufficient evidence to determine if the noise is associated with band noise, or noise associated with 50 cm observations with the CPSR2 backend. If we account for this noise, we can set a limit on the GWB in the three-frequency dataset that is comparable to limit we can set using only the 10 cm observations.

If we consider only the 10 cm and 20 cm observations, after accounting for the instrumental offsets, we find that there is no evidence for band or system noise (see Table S7). If we include the 20cm data, we find a 10% improvement in the limit with the PSR J1909–3744 data alone providing a 95% confidence limit on $A_{c,yr}$ of $9\times10^{-16}$. Techniques that account for this excess noise and properly account for DM variations are currently being developed.

## S2.2. PTA frequency response and PTA sensitivity curves

In order to calculate a limit on the fractional closure density of the Universe, $\Omega_{GW}$, we need to determine the frequency of mean sensitivity, $f_{GW}$. We calculated this frequency using three techniques that give consistent results.

The first technique has previously been applied to PTA data sets to determine the frequency of mean sensitivity (*8,29*) and enables limits on different power-law spectra to be derived from our result. Under the assumption that the limit is set in a narrow frequency range, and the characteristic strain spectrum has the form $h_c \propto (f_c)^\alpha$, the limiting strain amplitude, $A_{95}$, will be proportional to $(f_{GW})^\alpha$ (see right panel of Fig. S2). For our data set, we find good agreement with this scaling, and find $f_{GW} = 1/5.1$ yr$^{-1}$. This is shorter than the total data length because of the need to model both the pulsar spin down and varying TOA quality through the data set, with the earlier data having poorer timing precision. Our limit is at a higher frequency than a previous limit on the GWB obtained with PPTA data (square P13 in Figure 2), because we have excluded early 20 cm data that is of poorer quality and likely contains low frequency noise associated with dispersion-measure variations.

The sensitivity of the data sets can be also observed through analysis of the power-spectral density estimates of the data sets. In the left panel of Fig. S2, we show for the dominant pulsar J1909–3744 the sensitivity (defined to be the 1-σ upper limit) of the power spectral density (PSD) estimates (black curve) formed from a weighted Lomb-scargle periodogram that is calculated in the presence of the timing model. At low frequencies, the sensitivity is greatly degraded because these periods are covariant with parameters that model the pulsar spin period and period derivative. The frequency that is most sensitive to a power law GWB (displayed as the blue line) is at approximately $f_{GW}$.

In the third case, we use a model-independent method to search for a GWB (*15*) that enables us to place limits on the strain amplitude $h_c$ in a series of harmonically related frequencies. This is in contrast to other searches in which we constrain the strain-amplitude spectrum to follow a functional form (i.e., power-law for a purely GW-driven background). For our data set, the two lowest frequency channels were the most sensitive to the GWB, indicating that the frequency of greatest sensitivity is < 0.3 yr$^{-1}$.

In Fig. 2, we show the sensitivity curve generated using the first technique, converted to a characteristic strain spectrum, for our observations (black curve). To generate this curves, we converted the power-spectral density estimates to 2-σ false-alarm probabilities. We assumed that the PSD had 325 degrees of freedom (corresponding to the number of points in our data set), which is a reasonable assumption because our data are relatively uniformly spaced.

We also estimated the sensitivity of a potential observing campaign with the Square Kilometre Array Telescope (SKA, ref. *26*). We assume a timing campaign of 5.5 yr duration with the cadence achieved by the PPTA, but with nominal timing precision of ~ 12 ns uncertainty per observation. This uncertainty represents the single pulse jitter limit in 1~hr of observation estimated for this pulsar (*35*). The limit on the GWB obtained by this campaign is displayed as the blue pentagon in panel *d* of Figure 2.

## S3. Model Comparison

We compared our limit with the five most recent predictions for the GWB (*9-12, 22*). Each of the models predicts the amplitude with some uncertainty and presents the result as a probability density function $\rho_M(A_M)$. Each model also predicts a specific strain-spectral shape. We calculated the probability that these models agree with the data in two ways, which were found to give consistent results.

In the first technique, we used a Bayesian evidence-based method to calculate the probability that the predicted GWB was consistent with the data. Instead of placing a uniform or log-uniform prior on the amplitude $A_{c,yr}$, we used the predicted amplitude range given by a specific model, $\rho_M(A_M)$, as a physical prior for a given model when searching for a GWB in our data set. The probability-density functions (PDFs) $\rho_M(A_M)$ are specific to each model considered. We then compared the Bayesian evidence for these searches to that from an analysis that did not include a search for a GWB. The probability that the model is consistent with the data is then the ratio of the evidences, $P_{evid}$. These values are found in Table S2.

We also used a previously used technique to compare limits to models (*8*) that yields consistent results. Here, we assume that $A_M$ and $A_{lim}$ are independent random variables, drawn respectively from PDFs $\rho_M(A_M)$ and $\rho_{lim}(A_{lim})$. The PDF $\rho_M(A_M)$ describes a model prediction and $\rho_{lim}(A_{lim})$ is the posterior PDF for $A$ calculated from the data. For the two distributions to be consistent with each other we require $A_{lim} > A_M$. Therefore the probability of consistency between a model and the data is

$$P_{model}(A_{lim} > A_M) = P(A_{lim} - A_M > 0) = \int_0^\infty dA_M\, \rho_M(A_M) \int_{A_m}^\infty dA_{lim}\, \rho_{lim}(A_{lim}). \quad (S1)$$

If the second integral is evaluated, equation (S1) is the integral of the product of the PDF of the model and the CDF of the limit, and matches expressions previously used to compare limits to models (*8*).

For all of the models, the predicted values of $A_{c,yr}$ followed lognormal distributions:

$$\rho_M(A_{c,yr}) = (2\pi\sigma^2 A_{c,yr}^2)^{-1/2} \exp[-(\log(A_{c,yr}) - \mu)^2/2\sigma^2] \quad (S2)$$

where $\mu$ and $\sigma$ are, respectively, the mean and standard deviation of the distributions.

In Table S8, for each model we list both its parameters and the probability $P$ that it is in agreement with the limit.

In one case (*10*), the strain spectrum has been modeled to include a low-frequency cutoff due to both the coupling between binary SMBHs and their stellar environments, and stalled mergers of the SMBH binaries. The cutoff is modeled by adding a pre-factor to the strain spectrum of the form $\exp[-(f/f_c)^{-4}]$, where $f_c$ is a cutoff frequency and is modeled to be 0.20 yr$^{-1}$.

This prefactor modifies the shape of the power spectral density induced by the GWB to be

$$P_{\text{GW}}(f) = A_{c,yr}^2/(12\pi^2) \, (f/1 \text{ yr}^{-1})^{-2/3} \exp[-2(f/f_c)^{-4}]. \quad (S3)$$

Because this cut-off frequency is at least partially motivated by evolution driven by the scattering of environmental stars off binary SMBHs (a process we wish to consider independently) we calculate the probability that the model is in agreement with the limit both including and excluding the pre-factor. We assumed that the uncertainty for the power-law model applies to the exponential model. In both cases, we find the model is inconsistent with the data with high probability (see Table S8). Despite having a cut off at a frequency above our frequency of maximum sensitivity, we are able to exclude this model because it predicts a very strong signal at higher frequencies.

In another case (*22*), the effects of mergers driven by stellar environments result in the prediction of an altered strain spectrum, with two spectral breaks:

$$h_{c,\text{env}}(f) = A_{c,yr} f_2^{-2/3} / [\, (f/f_1)^{-2} + (f/f_2)^{-1/2} + (f/f_2)^{-4/3} \,]^{1/2}. \quad (S4)$$

The lowest spectral break occurs at $f_1 \approx 0.01 \text{ yr}^{-1}$ and corresponds to the frequency at which SMBH binaries start to decouple from their environments. The second occurs at $f_2 \approx 0.3 \text{ yr}^{-1}$, and defines the frequency at which decoupling completes. The power spectral density associated with this model is

$$P_{\text{GW}}(f) = h_{c,\text{env}}(f)^2 \, f^{-3}/(12\pi^2). \quad (S5)$$

## S4. Evolution of binary separation

Fig. 3 schematically presents various evolutionary scenarios of a system comprising two $10^9$ solar-mass SMBHs, following general arguments (*4*) that have been recently reviewed (*21*). We assume that the binaries are initially separated by $10^4$ pc. We further assume for the purposes of demonstration in Fig. 3 that the evolution depends only on the masses of the SMBHs and their surrounding stellar environment. In this scenario, the mechanism for solving this "final-parsec problem" is assumed to be three-body interactions between the binary and stars on radial orbits (*4, 47-48*). While some of the most massive binary SMBHs that likely dominate the GW signal are also expected to be gas-poor (i.e., "dry", Refs. *11, 12*), the role of gas in shrinking binary orbits could be crucial in other systems (*21, 24, 49-50*). In the following, we outline the physically motivated assumptions behind Fig. 3.

Through dynamical friction (*51*), the binaries are driven to coalescence at a rate given by

$$\dot{a}_{\text{DF}}/a = 5 \times 10^{-6} \text{ yr}^{-1} \, \log(N_\star) \, (M_{\text{tot}}/2M_9) \, (a/100 \text{ pc})^{-2} \, (200 \text{ km s}^{-1}/\sigma_\star), \quad (S6)$$

where $a$ is the pair separation, $\dot{a}_{\text{DF}}$ is its first time derivative corresponding to the effects of dynamical friction, $M_{\text{tot}}$ is the total mass of the SMBHs, $M_9$ is the $10^9$ solar masses, $N_\star$ is the number of stars comprising the environment and $\sigma_\star$ is the velocity dispersion of the stars. We assume that the environment comprises $2\times10^8$ stars and that $\sigma_\star = 200$ km s$^{-1}$ in the fiducial case.

The SMBHs will form a bound system when the stars enclosed in the orbit have mass comparable to the mass of the more massive SMBH ($M_1$). This occurs when the separation between the SMBHs is equivalent to the gravitational influence radius of the more massive SMBH. For a simple stellar density model (*52*) of a cusp with power-law index $\gamma$ within the gravitational influence radius of the more massive SMBH and an isothermal sphere beyond the influence radius with velocity dispersion $\sigma_\star$, the influence radius is given by

$$r_{\text{inf}} = (3 - \gamma)\, G M_1 / \sigma_\star, \qquad\qquad (S7)$$

where $G$ is the universal gravitational constant. Relating $M_1$ with $\sigma_\star$ using the empirical relation between SMBH masses and the velocity dispersions of galaxy bulges (*1*), we obtain the expression presented in the main text for a typical formation separation of a binary SMBH ($a_{\text{form}} = 60 M_9^{0.54}$ pc).

When the binary velocity becomes comparable to the velocity dispersion of the surrounding stars, dynamical friction becomes ineffective at guiding the merger of the system. This occurs at

$$a_{\text{hard}} = 50\text{ pc } q/(1+q)^2\, (M_{\text{tot}} / 2\, M_9)\, (\sigma_\star / 200\text{ km s}^{-1})^{-2}, \qquad\qquad (S8)$$

where $q$ is the mass ratio of binaries.

At this point the effects of dynamical friction are greatly inhibited, but the system can still be driven to smaller separation through the scattering of stars. Circumbinary gas can play a similar role. In this regime, the orbital separation evolves as

$$\dot{a}_{\text{hard}}/a = 7\times10^{-6}\text{ yr}^{-1}\, (200\text{ km s}^{-1}/\sigma_\star)\, (10^4\, M_{\text{sun}}\text{ pc}^{-3} / \rho_\star)\, (a/\text{pc}), \qquad\qquad (S9)$$

where $\rho_\star$ is the stellar density and is assumed to be $10^4\, M_{\text{sun}}$ pc$^{-3}$ in the fiducial case, and $M_{\text{sun}}$ is one solar mass.

In the final stages, the system is driven through GW emission:

$$\dot{a}_{\text{GW}}/a = 1.5\times10^{-4}\text{ yr}^{-1}\, q/(1+q)^2\, (a/\, 0.1\text{ pc})^{-4}\, (M_{\text{tot}}/ 2\, M_9)^3. \qquad\qquad (S10)$$

Equation (S10) applies only to binaries in circular orbits. We note that large orbital eccentricities, potentially incited by binary environments (*52*), redistribute the spectral-energy distribution of the GWB from lower to higher frequencies, further enhancing the low-frequency attenuation of the GWB caused by the environments (*22*). For a range of realistic distributions of binary eccentricities, the attenuation of the GWB signal due to eccentricities is subdominant to effects of the environments themselves in accelerating binary evolution, regardless of the orbits (*22*).

The curves in Fig. 3 were constructed as follows:

*Fiducial case (blue curve):* We assume that dynamical friction and stellar-driven binary hardening occur according to Equations (S6) and (S9) respectively.

*Slow merger (green curve):* We assume that dynamical friction (Equation S6) is a factor of $10^{-4}$ weaker than the fiducial case. As a result the, SMBHs do not form a binary as rapidly as in the fiducial case and may not merge within a Hubble time. This is also representative of the case where the galaxy merger timescale has been overestimated, which similarly lengthens the time it takes SMBH binaries to form gravitationally radiating binaries.

*Strong binary environment (gray curve):* We assume that, after the SMBHs form a binary that hardening (Equation S9) is a factor of $10^2$ greater than the fiducial case. As a result, the SMBH binary inspiral more quickly, and that environment drives inspirals to smaller separations.

*Stalled binary SMBH (red curve):* We assume that, after the SMBHs form a binary, inspiral due to hardening is $10^{-2}$ weaker than the fiducial case. As a result, binary inspiral does not occur within the age of the Universe.

*SMBH recoil (pink curve):* The purple-dashed curve shows the trajectory of the coalesced SMBH, relative to the centre of the galaxy, if a kick of 500 km s$^{-1}$ (*53, 54*) is applied after the merger. The SMBH becomes gravitationally unbound from the system because this is in excess of the escape velocity of the galaxy.

In Fig. 3, we also highlight (in light blue and pink respectively) the portions of the *'Fiducial'* and *'Strong environment'* curves where $\dot{a}_{GW} > \dot{a}_{hard}$, which is where GW emission dominates the energy loss.

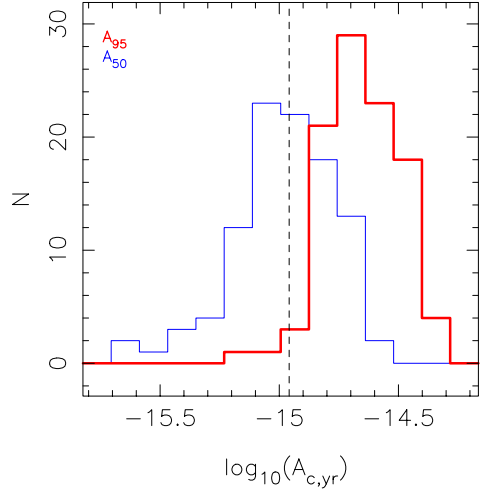

**Fig. S1:** Histogram of 95% (red thick histogram) and 50% confidence limits (blue thin histogram) for 100 simulated data sets with identical cadence and white noise to our PSR J1909–3744 observations, but with an injected GWB with $A_{c,inj}=1.1\times10^{-15}$ (dashed black vertical line).

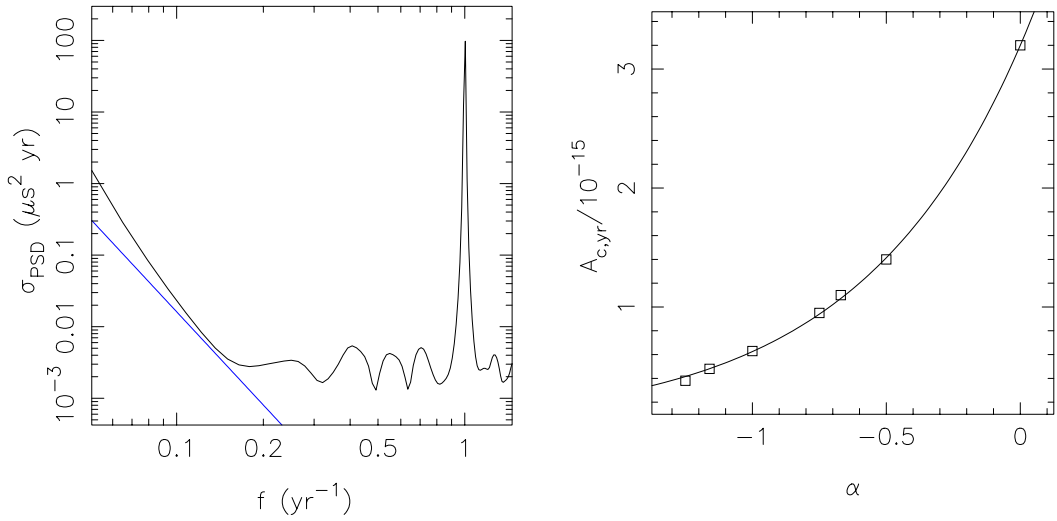

**Fig. S2:** Left panel: Sensitivity of power spectral density estimates of PSR J1909–3744 (black curve). The blue line shows the power spectral density of a GWB, showing that the data set is most sensitive to a GWB at frequencies of 0.2 yr$^{-1}$. The peak at 1 yr$^{-1}$ is associated with fitting for pulsar position and proper motion. Right panel: Limit on the GWB for power-law strain spectra $h_c(f) = A_c\,(f/\,1\text{ yr}^{-1})^\alpha$ set at different $\alpha$ (open boxes), and the best-fitting curve (black curve).

|  | **Δ log (E)** | | | |
| --- | --- | --- | --- | --- |
| **Noise Model** | J1909–3744 | J1713+0747 | J1744–1134 | J0437–4715 |
| White | **0** | 0 | **0** | 0 |
| White + Red | -0.2 | **6.1** | -0.3 | **329.1** |
| White + GWB | 0.1 | -0.7 | 0.0 | 324.6 |
| White + Red + GWB | -0.5 | 5.4 | -0.7 | 326.9 |
| $A_{c,yr,95}/10^{-15}$ | **1.0** | **3.9** | **10.6** | **7.4** |

**Table S1:** Evidence comparison for pulsar timing models, relative to a white noise only model. Evidence is measured relative to models containing only excess white noise. The favoured models, selected because they have the largest evidence, are displayed in bold typeface. We also list the 95% confidence limits on the GWB obtained from individual pulsars in our sample.

| Right ascension (hh:mm:ss): | 04:37:15.8961747(6) |
|---|---|
| Declination (dd:mm:ss) | −47:15:09.11071(6) |
| Pulse frequency ($s^{-1}$): | 173.6879458121831(3) |
| First derivative of pulse frequency ($s^{-2}$): | −1.728359(5) ×$10^{-15}$ |
| Dispersion measure ($cm^{-3}$ pc) | 2.648(5) |
| Proper motion in right ascension (mas $yr^{-1}$): | 121.4401(16) |
| Proper motion in declination (mas $yr^{-1}$): | −71.4732(17) |
| Parallax (mas): | 6.43(7) |
| Orbital period (d): | 5.7410460(6) |
| Epoch of periastron (MJD): | 54530.1727(5) |
| Projected semi-major axis of orbit (lt-s): | 3.36671473(11) |
| Longitude of periastron (deg.): | 1.39(3) |
| Orbital eccentricity: | 1.91820(19) ×$10^{-5}$ |
| First derivative of orbital period: | 3.721(12) ×$10^{-12}$ |
| Periastron advance (deg. $yr^{-1}$): | 0.014(3) |
| Companion mass ($M_\odot$): | 0.211(19) |
| Longitude of ascending node (deg.): | 205.8(15) |
| Orbital inclination angle (deg.): | 137.3(10) |
| Epoch (MJD): | 54500 |

**Table S2:** Ephemeris for PSR J0437−4715. For each parameter, the 1-σ uncertainties in the last digit(s) are listed in parentheses.

| Right ascension (hh:mm:ss): | 17:13:49.5327232(16) |
|---|---|
| Declination (dd:mm:ss): | +07:47:37.49790(5) |
| Pulse frequency ($s^{-1}$): | 218.8118404347997(5) |
| First derivative of pulse frequency ($s^{-2}$): | −4.08383(5)×$10^{-16}$ |
| Dispersion measure, DM ($cm^{-3}$ pc): | 16.003(8) |
| Proper motion in right ascension (mas $yr^{-1}$): | 4.925(7) |
| Proper motion in declination (mas $yr^{-1}$): | −3.928(13) |
| Parallax (mas): | 0.75(9) |
| Orbital period (d): | 67.82515(3) |
| Epoch of periastron (MJD): | 51997.5785(15) |
| Projected semi-major axis of orbit (lt-s): | 32.3424217(5) |
| Longitude of periastron (deg.): | 176.191(8) |
| Orbital eccentricity: | 7.49408(20)×$10^{-5}$ |
| Companion mass ($M_\odot$): | 0.27(4) |
| Longitude of ascending node, (deg.) | 88(7) |
| Orbital inclination angle (deg.) | 74(2) |
| Epoch: | 54500 |

**Table S3:** Ephemeris for PSR J1713+0747. For each parameter, the 1-σ uncertainties in the last digit(s) are listed in parentheses.

| Right ascension (hh:mm:ss): | 17:44:29.405794(4) |
| --- | --- |
| Declination (dd:mm:ss): | 11:34:54.6809(3) |
| Pulse frequency ($s^{-1}$): | 245.4261197130577(9) |
| First derivative of pulse frequency, ($s^{-2}$): | $-5.38158(7) \times 10^{-16}$ |
| Dispersion measure ($cm^{-3}$ pc): | 3.131(16) |
| Proper motion in right ascension (mas $yr^{-1}$): | 18.782(12) |
| Proper motion in declination (mas $yr^{-1}$): | $-9.47(6)$ |
| Parallax (mas): | 2.7(1) |
| Epoch: | 54500 |

**Table S4:** Ephemeris for PSR J1744−1134. For each parameter, the 1-σ uncertainties in the last digit(s) are listed in parentheses.

| Right ascension: | 19:09:47.4346740(6) |
| --- | --- |
| Declination: | $-37$:44:14.46670(3) |
| Pulse frequency ($s^{-1}$): | 339.31568728824556(16) |
| First derivative of pulse frequency ($s^{-2}$): | $-1.6148237(12) \times 10^{-15}$ |
| Dispersion measure ($cm^{-3}$ pc): | 10.3955(19) |
| Proper motion in right ascension (mas $yr^{-1}$): | $-9.5204(19)$ |
| Proper motion in declination (mas $yr^{-1}$): | $-35.768(7)$ |
| Parallax (mas): | 0.855(18) |
| Orbital period (d): | 1.5334494747(3) |
| Epoch of periastron (MJD): | 53631.47(4) |
| Projected semi-major axis of orbit (lt-s): | 1.89799117(5) |
| Longitude of periastron (deg.): | 176(9) |
| Orbital eccentricity: | $9(1) \times 10^{-8}$ |
| First derivative of orbital period: | $5.02(5) \times 10^{-13}$ |
| Companion mass ($M_\odot$): | 0.209(3) |
| Longitude of ascending node (deg.): | 150(100) |
| Orbital inclination angle (deg.): | 93.44(11) |
| Epoch (MJD): | 54500 |

**Table S5:** Ephemeris for PSR J1909−3744. For each parameter, the 1-σ uncertainties in the last digit(s) are listed in parentheses.

| Model | $\Delta\log(E)$ | $A_{95}/10^{-15}$ |
|---|---|---|
| Red + DM | 0 | 2 |
| Red + DM + Group Noise | 3.8 | 1 |
| Red + DM + Band Noise | 4.5 | 1 |

**Table S6:** Evidence comparison for three-band dataset for PSR J1909–3744. A significant (P > 98%) increase in evidence is found by including extra noise terms. There is insufficient change in evidence to distinguish band noise from group noise.

| Model | $\Delta\log(E)$ | $A_{c,yr,95}/10^{-15}$ |
|---|---|---|
| Red+DM | 0 | 0.9 |
| Red + DM + Group Noise | -1.4 | 0.9 |
| Red + DM + Band Noise | -0.5 | 0.9 |

**Table S7:** Evidence for comparison for two-band analysis for PSR J1909–3744 using only 20 cm and 10cm observations. Models including extra noise processes are disfavoured because the evidence decreases.

| Model | $\mu$ | $\sigma$ | $P_{model}$ | $\Delta\log(E)$ | $P_{evid}$ |
|---|---|---|---|---|---|
| S13 | -14.96 | 0.28 | 0.09 | -2.3 | 0.09 |
| R14[a] | -14.90 | 0.40 | 0.21 | -1.7 | 0.20 |
| R15 | -14.95 | 0.18 | 0.06 | -3.0 | 0.05 |
| M14 | -14.39 | 0.26 | 0.002 | -5.7 | 0.003 |
| Exp[b] | -14.39 | 0.26 | 0.04 | -2.7 | 0.06 |
| K15 | -14.70 | 0.10 | 0.005 | -5.3 | 0.005 |

*Notes: (a) Broken power-law model defined in Equation (4) (b) M14 model, including an exponential cut-off.*

**Table S8:** Predicted amplitudes of the GWB for models for the GWB (*9-12, 8*), and the probabilities $P_{model}$ that they are consistent with the limit derived from our observations. We show the probability calculated using equation (S1), labeled $P_{model}$, and using the difference in the logarithm of the evidence $\Delta\log(E)$ between each model and a model that does not contain a GWB, labeled $P_{evid}$.